\documentclass[twoside,a4paper]{article}
\usepackage[T2A]{fontenc}
\usepackage[cp1251]{inputenc}
\usepackage[english]{babel}
\usepackage{graphicx}
\makeatletter
\renewcommand{\@cite}[2]{{#1 \if@tempswa   #2\fi}}
\renewcommand{\@biblabel}[1]{\hfill}
\makeatother

\newcommand{\beq}{\begin{equation}}
\newcommand{\eeq}{\end{equation}}
\newcommand{\bea}{\begin{eqnarray}}
\newcommand{\eea}{\end{eqnarray}}

\def\sun{\hbox{$_\odot$}}

\begin{document}

\vbox{}

\begin{center}
{\LARGE  \bf Kinematics of SS\,433 radio jets: interaction of jets with ambient medium}
\vspace{0.7cm}

{\large  A.~A. Panferov}

IPT, Togliatti State University, Russia

E-mail: panfS@yandex.ru
\end{center}
\vspace{0.7cm}

\begin{abstract}
The problem of the interaction between the precessing mildly relativistic
radio jets of SS\,433 and an ambient medium has been often invoked in interpretations
of the jets behaviour. The dynamical interaction 
could be responsible for disagreements between the kinematic model and observations
of the jets. To estimate the relative ram pressure of the jet and impinging matter 
we used the profile along the radio jets of brightness of synchrotron radiation.
From this we estimated magnitude of possible deceleration and twisting of the jets,
and compare model locus of the jets with observations.
The magnitude is enough big to be observed, that addresses an applicability
of the kinematic model, based mostly on the
optical observations, to the radio jets and
a problem of the kinematic distance to the object in particular.

\end{abstract}

\section*{\large Deviation from kinematic model}
Jets of the X-ray binary star SS\,433, the prototype of microquasars, are best studied.
The kinematic model (e.g., \cite[1981]{HJ81})
was developed mainly on the base of data on the
optical jets (extended to distances of $\sim 10^{15}$\,cm), and assumes the
independent of each other blobs in the 
SS\,433 precessing jets move at a constant velocity (rectilineally  and uniformly).
This model is a basement for any physical model of the jets. 
However, observations evidence about deviation of the radio jets from
the kinematic model by more than 10 percents
(e.g., \cite[2004]{Sch04}; \cite[2008]{Rob08}), 
in particular about regular shift in the precession phase 
(\cite[2004]{St04}) --- evidently that the optical and radio jets are
some decoupled in kinematics. Moreover, the published kinematic modelings of morphology of the
radio jets at distances from source of $10^{16}$--$10^{17}$\,cm (e.g., \cite[1981]{HJ81}) 
might be accommodated to a jet velocity and a distance to the object essentially
varied simultaneously around those of the canonical kinematic model 
$v_{\rm j}=0.26\,c$ and $D=5.5$~kpc, respectively, where $c$ is the speed of light.

\section*{\large Model of dynamics of radio jets}
Powerful wind from supercritical accretion disk in SS\,433 outflows at a rate
of $\sim 10^{-4}$~M\sun/yr with the velocity $v_{\rm w} \sim 1500$~km/s. 
In the model of \cite[(2006)]{Beg06}
this wind is able even to transmit the precession and nutation rotations of the
disk to the jets near by the source. Therefore the jets may decelerate
while a movement through this wind.

The polarization observations of \cite[(2008)]{Rob08} reveal that the jets magnetic field
is aligned with the spiral of the precessing jets, and not with the jet velocity
vector (also see \cite[2004]{St04}), and the apparent at 15 GHz leading edge of the jets is
the most bright and sharp. This suggests a significant interaction of  
the jets with a surrounding medium, influencing on morphology of
the jets. In particular, this interaction might affect jets dynamics. 

%
\begin{figure}[ht]
\centerline{\hbox{\includegraphics[width=7.5cm]{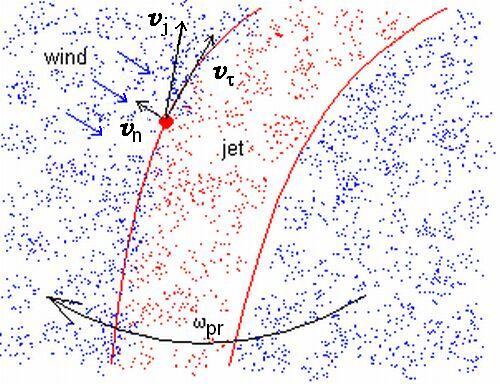}}}
\caption{
The sketch of precessing jet. The relative ram pressure of the jet and wind
is determined by the component of jet velocity normal to
jet-ambient medium interface.
}
\end{figure}
%

The sketch of precessing jet ploughing through ambient medium is shown in Fig.~1. 
Our model of the dynamical interaction of precessing jets with ambient medium
supposes the followings: the surrounding matter at rest as $v_{\rm w} \ll v_{\rm j}$; 
the jet as a rather continuous flow, whose
local velocity vector is somehow inclined to the jet precession spiral;
in the jet co-moving reference frame the surrounding gas impinges the jet front 
surface, not penetrating
into, with the resulting ram pressure $p_{\rm dyn}=\rho_{\rm a} v_{\rm n}^2$, 
where $\rho_{\rm a}$ is the density of the surrounding gas, 
$v_{\rm n} = v_{\rm j}(1+1/(\omega_{\rm pr} t \sin \theta_{\rm pr})^2)^{-1/2}$ is the
component of the jet velocity normal to the surface,
$\omega_{\rm pr}$ and $\theta_{\rm pr}$ are the precession frequency and 
the opening half-angle of the precession cone, respectively, $t$ is the jet flight time.

%
\begin{figure}[hb]
\centerline{\hbox{\includegraphics[width=11cm]{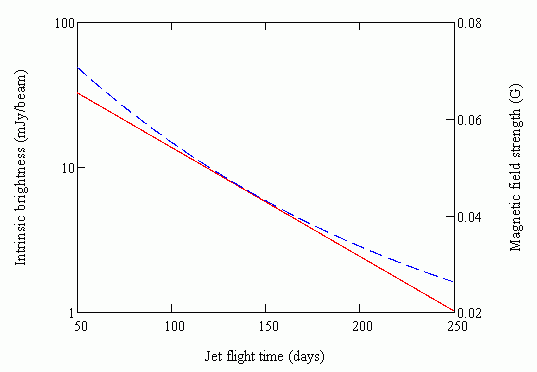}}}
\caption{
Intrinsic brightness (solid line) at 4.86~GHz and the equipartited magnetic field 
(dashed line) of the jets of SS\,433 in dependency on the jet flight time, for
parameters of the model given in the text.
}
\end{figure}
%

As the jets propagate through the slow wind the shock waves develop
at the jets surface, providing energetics of synchrotron radiation of the jets.
The minimum pressure in the jets --- the lower limit of the jets 
internal pressure $p$ --- can be evaluated from brightness of the radiation
in an approximation of energy equipartition between relativistic particles and
magnetic field, as $p_{\rm min}=H^2/4\pi=A(\beta_{\rm e} L_{\rm r}/f V)^{4/7}$,
where the dependency on the index $\alpha$ and frequency
range of the synchrotron spectrum is involved in the coefficient $A$,
$L_{\rm r}$ is the radio luminosity in this range of an optically thin jet region 
of the volume $V$, $f$ is the filling factor of the region, 
$\beta_{\rm e}$ is the ratio of the total particle energy
density to the electron energy density,
$H$ is the strength of the equipartited magnetic field (\cite[1970]{Pac70}).
The profiles of intrinsic brightness (in a jet co-moving reference frame) of the jets 
in SS\,433 at 4.86~GHz and derived from it magnitude of the equipartited
magnetic field in the jets are shown in Fig.~2 as functions of the jet flight time. 
The brightness is adopted from the fit of \cite[(2010)]{Rob10}
over 60 -- 150 days of jet age, as $B=32.7\exp(-(t-50)/57.7)$~mJy/beam, where the
jet age $t$ is in units of days. The dependency $B(t)$ is extrapolated
beyond the definition region. This simplification is justified by the
posteriority that most of jet deceleration takes place in this region (see below).
The magnetic field is derived for $\alpha = 0.7$ 
(for $B\propto \nu^{-\alpha}$), $\beta_{\rm e}=100$ and $f = 10^{-3}$.

Above mentioned data suggests the front side of the jets undergoes an
essential ram pressure of the impinging gas, therefore the internal jet pressure
should approximately equal the dynamical pressure at the jet surface: 
$p_{\rm dyn} \approx p \geq p_{\rm min}$.
So, the model allows to evaluate the dynamical pressure on the jets,
which the surrounding matter renders, from minimum energy of the relativistic
particles and the magnetic field in the jets. To find space disposal of the jets, i.e.
the overall jets kinematics, the kinematics of each enough small segment of the jets 
was calculated,
taking into account an action of the dynamical pressure on jets movement. In this
procedure the segments were supposed independent of each other. Initial conditions
of the kinematic task were determined by the canonical kinematic model of
the jets with the parameters collected in \cite[(2010)]{Pan10}.

\section*{\large Manifestation of dynamical interaction from modeling}
1. Estimation of the ram pressure on the jets, from brightness of the jets
synchrotron radiation, shows that density of the wind falls 
as $t^{-2}$ in the region of most of jet deceleration $t\le 260$~days (see below)
and is approximately two order less than could be expected in the case of
the isotropical wind from supercritical accretion disk in SS\,433 with
an outflow rate of $10^{-4}$~M$_\odot$/yr.
%
\begin{figure}[ht]
\centerline{\hbox{\includegraphics[width=11cm]{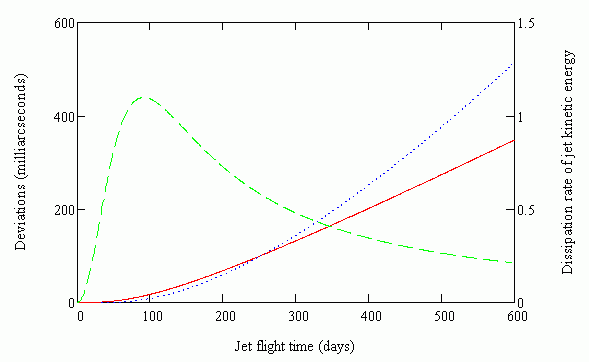}}}
\caption{
Transverse (solid line) and longitudinal (dotted line) deviations
of the kinematics of a jet of SS\,433 from free ballistic movement,
represented by the canonical kinematic model, as functions of the jet
flight time, laid out along the left axis. 
And dissipation rate of the kinetic energy of a jet
(dashed line), laid out along the right axis, in units of a thousands of
initial kinetic energy flux of the jet per a day. The plots are calculated
for the wind ram pressure corresponding to the magnetic field in Fig.~2.
}
\end{figure}
%

2. Deviations from the kinematics of a free ballistic movement prescribed 
by the canonical kinematic model, across and along the jets,
are calculated with described above dynamical model. They are shown in Fig.~3 as functions
of the flight time, for the ram pressure corresponding to the magnetic field in Fig.~2.

These deviations can also be viewed in Fig.~4 as a difference between the tracks
of the canonical kinematic model and the dynamical model. The tracks are
overlaid on the image of the jets, taken from \cite[(2004)]{BB04}.

The dynamical model for a filling factor of a synchrotron
radiation region of $\sim 10^{-3}$ is in compliance with
the observed deviations magnitude, $\sim 10\%$: 
the deviations developed in the precessing jets flowing through the medium
can be as much as several 100 mas at the jet length of $\sim 4''$.
It seems from Fig.~4 the brightness ridge of the jets lies somewhere between
models for $f = 10^{-3}$ and $10^{-2}$.

%
\begin{figure}[ht]
\centerline{\hbox{\includegraphics[width=13cm]{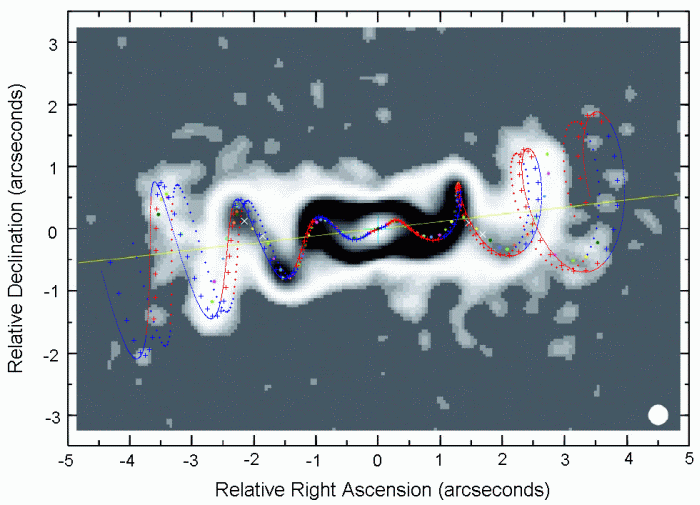}}}
\caption{
Model tracks of the precessing jets of SS\,433 superposed on the VLA image 
from \cite[(2004)]{BB04}. The jet brightness ridge is marked by colored points. 
The solid spiral is 
the canonical kinematic model, without accounting for the jet-wind dynamical interaction,
the dotted spiral, with dots at 5 day intervals, the model with the accounting, assuming
the filling factor of a synchrotron radiation region $f=10^{-3}$, and the crossed
spiral, with crosses at 10 day intervals, the model with $f=10^{-2}$. 
The blue and red colors of the spirals indicate approaching 
and receding regions in the jets, respectively. The length of a jet track is 560 flight days.
}
\end{figure}
%

3. There is also shown in Fig.~3 the distribution along the jets of 
dissipation rate of the kinetic energy of the jets of SS\,433. Most of the dissipation,
and of the jets deceleration, takes place in the region 30--260 flight days,
and the integrated dissipation rate achieves 10\% of the jets kinetic luminosity 
in the first 130 days. 

It should be noted, the energy lost by jets does not go into the jets radiation: 
the radio luminosity is approximately $10^{-4}$ times the
kinetic luminosity, of the same order is X-ray luminosity
of the jets extended to the bounds of W\,50 (\cite[2006]{Bri07}). 
However, this contradiction does not restrict the dynamical model because
the dissipated energy could be redirected into mechanical energy of 
the surrounding medium. Indeed, observations evidence that most of the power of 
the jets is not radiated and is somehow transfered into mechanical energy of W\,50
and the surrounding interstellar matter (e.g. \cite[1998]{Dub98}).

\section*{\large Conclusions}
None models in Fig.~4 fit acceptably good the jets ridge.
Nevertheless, as we have shown the deceleration and azimuthal bending (around the
precession axis) of the jets of SS\,433 are enough big and should be 
accounted in interpretation
of the observations, which evidence as we know about an essential  
deviation of the radio jets from the kinematic model. In particular
this might affect the kinematic distance to SS\,433.
In its estimation one usually assumes that velocity of the radio jets
equals the value derived for the optical jets, i.e. for the jets region close 
to the source, that could be rather crude approximation (however, see 
another approach of \cite[2004]{BB04}).
By different estimations the distance to SS\,433 is in the range 3--5.5~kpc.
Its improving by the kinematic method is an important task.
The studying of the jets deceleration needs observations of jets
regions as far from the
source as possible, where the dynamical jets-wind interaction develops most
strongly in jets morphology. Besides, it would be more appropriate to fit
the kinematic model track to the overall jet morphology, and not to the bright ridge,
because the latter could lie not along the axis of the precessing jets 
(\cite[2003]{Al03}; \cite[2008]{Rob08}).

\end{document}